# Old wine in new bottles

## – From Onsager relations to spincaloritronics –

**Klaus Stierstadt**

Fakultät für Physik der Universität München (LMU), Schellingstrasse 4, 80799 München, Germany

Onsager's reciprocity relations for the coefficients of transport equations are now 87 years old. Sometimes these relations are called the Fourth Law of Thermodynamics. Among others they provide an effective criterion for the existence of local equilibrium and of microscopic reversibility. Since the beginning of our century Onsager's relations have seen a revival in the field of spincaloritronics. There the relations are very helpful in judging the utility of modern devices for electronic data processing.

## I. Introduction

All kinds of motion we observe are of *irreversible* nature. In thermodynamic language this means there exists always some change in the surroundings of the moving object. But today teaching thermodynamics on the undergraduate level is mainly based on equilibrium states and on *reversible* processes alone. This contradiction is one of the causes which make thermodynamics so challenging for our students.

We deal in this article with one of the corner-stones of irreversibility: The so called Fourth Law of Thermodynamics, that means Onsager's reciprocity relations for transport coefficients. And they are of major relevance for modern electronic data processing. With Onsager's relations one may obtain a feeling why obvious different flows or fluxes are intimately interrelated by the behaviour of transport carriers, namely the atoms, electrons, phonons and so on. In the first half of this article Onsager's relations are explained. In the second half their bearing for spincaloritronics is delt with.

## II. Linear transport equations

*Panta rhei* stated Heraklitos, everything flows, and this is so in nature and in technology. You may think of the cosmos, of the wheather, the insides of plants and animals, as well as on water and electric currents, on pipelines, on communication etc. In the course of time physicists have discovered simple laws for some of these transport processes: Ohm's law, Fourier's law of heat conduction. Fick's law of diffusion and so on. The corresponding transport equations had been generalized by Lars Onsager (1903 – 1976) in the following way:

$$\boldsymbol{J}_X = L_{XY}\hat{\boldsymbol{F}}_Y. \tag{1}$$

Here $\boldsymbol{J}_X$ is he flow density or the flux of an extensive property $X$, $\hat{\boldsymbol{F}}_Y$ is a generalized force, or the driving force, and $L_{XY}$ is the kinetic coefficient (Fig. 1). This coefficient is simply related to the so called conventional transport coefficient as there are electrial conductivity, heat conductivity, diffusion constant etc. (examples in Appendix A).

A survey of the large number of known transport processes is best represented in matrix form as in Table 1. Here are listed many processes which are already observed and some others which are supposed to occur. In the first column of the table the fluxes $\boldsymbol{J}_X$ are shown for heat, volume, particles, electric charge and magnetic moment. The first row shows the corresponding driving forces $\hat{\boldsymbol{F}}_Y$ of temperature, pressure, chemical and electrical potential and of the magnetic field. These driving forces have been defined by Onsager in such a way to give a positive production of entropy for the process in consideration:

$$\boldsymbol{J}_X \bullet \hat{\boldsymbol{F}}_Y = \sigma \tag{2}$$

(see Appendix A for details). Here $\sigma$ is the entropy produced per unit volume and per unit time by the process in question, $\sigma = S/(Vt)$. The second diagonal in Table 1 is occupied by the so called 'conjugated' fluxes and forces which are characterized by the product $XY$ having unit of energy.

In principle all the effects shown in Table 1 can occur in every state of aggregation, in gases, liquids and solids. It is necessary of course that the substance contains mobile carriers of energy, momentum, electric charge etc., namely atoms, electrons, phonons, magnons and so on. If the carriers are of atomic size the process may be called osmosis, for larger particles the term kinesis is in use. In many cases the fluxes are much larger at the interface between two different materials ('heterogenous effects') than in a homogenous substance. But in principle all the processes can occur in every homogeneous material. In this case the effects are sometimes quite small and the experiments are difficult and expensive [27].

## III. The Onsager matrix

It is well known since long ago that one and the same flux can be produced by different driving forces which can also work simultanously. For instance the molecular energy of heat can be transported by a difference in temperature and one of the chemical or electrical potential (heat conduction, Dufour effect, Peltier effect). In such cases equation (1) must be extended in the following way

$$J_Q = L_{QT}\hat{F}_T + L_{Q\mu}\hat{F}_\mu + L_{Q\phi}\hat{F}_\phi \qquad (3)$$

(symbols as in Table 1). If we add still other fluxes we have

$$J_1 = L_{11}\hat{F}_1 + L_{12}\hat{F}_2 + L_{13}\hat{F}_3 + ...$$

$$J_2 = L_{21}\hat{F}_1 + L_{22}\hat{F}_2 + L_{23}\hat{F}_3 + ... \qquad (4)$$

$$J_3 = L_{31}\hat{F}_1 + L_{32}\hat{F}_2 + L_{33}\hat{F}_3 + ...$$

Here the kinetic coefficients $L_{ik}$ make up a matrix.

And now comes Onsager's great discovery in 1931 [1]: *This matrix is symmetric*, and so we have

$$L_{ik} = L_{ki}. \qquad (5)$$

Therefore we may infer from Table 1: The coefficient of thermodiffusion (Soret effect) is equal to that of the diffusion-caloric process (Dufour effect). And the coefficient of the Seebeck effect is equal to that of the Peltier effect etc. These connections are called 'Onsager's reciprocity relations'. For his discovery Onsager was awarded the Nobel price for chemistry in 1968: *'For the discovery of reciprocal relations bearing his name, which are fundamental for the thermodynamics of irreversible processes'*. Onsager found the unversally valid relations (5) by clever averaging the fluxes of extensive variables *X* subject to microscopic reversibility. This is explained in many textbooks (i. e. [2] – [6]) and a short survey is given in Appendix B. Because of its universal importance some authors call equation (5) the 'Fourth Law of Thermodynamics'.

## IV.  Local equilibrium

Of course equations of linerar transport like (1) are a mere approximation to the real world. They are the first term of a Taylor series valid for 'small' driving forces. If these forces become too large dramatic changes of the transport modes occur: electrical dicharge, turbulent flow, melting or evaporation of the substance etc. But what does mean 'small' in this context? Following Onsager 'small' means that local equilibrium is conserved. Then the laws of equilibrium thermodynamics should be valid in every small region of the sample. But this condition is essentially the same as microscopic reversibility: If the driving force is reversed the transport of porperty *X* must occur backwards in such a way that no noticeable changement can be observed in the surroundings. Consequently the meaning of 'small' is a question of the accuracy of measurement or of computation. And one may test this condition by letting the driving forces go to zero.

In this respect Onsager's relations are of special interest: If one has shown by experiment or by calculation that equation (5) is valid, then one can be sure to be inside the linear domain of equation (1). And then you can also use equation (3) and (4) and you can obtain the produced entropy by formula (2). Until now nobody has found a better test than Onsager's relations for the existence of local equilibrium.

Strictly speaking microscopic reversibility is equivalent to time reversal of the process in consideration. Therefore Onsager's relations (5) are not obeyed in the presence of a magnetic field. Reversing the field inverts the flux and instead of (5) one has

$$L_{ik}(B) = -L_{ki}(B). \qquad (6)$$

This relation had been deduced by Hendrik Casimir in 1945 [7].

## V. Magnetotransport effects

Since a long time it is known that many transport processes are influenced by a magnetic field. This gives rise to thermomagnetic and galvanomagnetic phenomena. They are well studied in the case of analogous Seebeck and Peltier effects. A magnetic field introduces additional gradients of temperature and electric potential which can point into different directions regarding the primary transport. A survey of the observed phenomena is shown in Figure 2. The transversal effects are caused primarily by the Lorentz force on moving electrons. The longitudinal ones are due to spatial variations of the electron density.

## VI. Experimental observations of Onsager's reciprocity

During thirty years following Onsager's discovery there appeared a large number of publications on experimental tests of the reciprocity. Most of them are theoretical concerning calculations of kinetic coefficients on microscopic base (e. g. [4] [8] [9]). Experimantal investigations were quite scarce (Fig. 3). They appeared mainly in the physico-chemical literature, probable because Onsager was awarded the Nobel price in chemistry. Before soon one was aware of the necessity to work very clean and carefully. The smallest impurities in the samples and the smallest differences of the apparatus were responsible for large dviations of the results. And there was an additional reason for the small number of reliable experiments: The 'inverse' effects, namely mirror inverted to the second diagonal in Table 1, are mostly very different in size. For instance the system iron-constantan has a large Seebeck but a very small Peltier effect. On the other hand in bismut-tellurid the reverse is found, very small Seebeck and large Peltier effect. Of course it is necessary to test Onsager's relations at one and the same substance.

A comprehensive review of the experimental work until 1960 was given by Miller [10]. It covers about 100 experimental papers from the upper left of Table 1. Most of them were able to show the validity of the reciprocity within an error of some percent. The deviations became much larger if the experiment was done with different samples of the same material or if the linear region was exceeded. Measurements of the Soret and Dufour effect are missing from Miller's survey. The first reliable results on these processes appeared not earlier than 1978 [11], carried out on the system $CCl_4$-$C_6H_{12}$ at room temperature with the result $L_{Q\mu}/L_{NT}$ = 1.00 $\pm$ 0,15.

During half a century following Miller's survey the interest in Onsager's relations ceased considerably. There appeared a number of investigations on liquid mixtures. A rather new publication [19] on molecular diffusion in mixtures of water and alcohol showed very good aggreement between $L_{N_1\mu_2}$ and $L_{N_2\mu_1}$. And in a recent experiment on thin films of nickel, permalloy or iron [18] the authors obtained perfect coincidence of Seebeck and Peltier coefficients at all temperatures between 77 and 325 K.

## VII. Spincaloritronics

Towards the end of the preceeding century suddenly arose a revival of Onsager's relation in the new field of spin transport. A pioneering work appeared in 1987 by Johnson and Silsbee [12] in which Table 1 had been completed with its sixth row and sixth column. These contain the flux $J_M$ of magnetic moments ('spins') and the driving force $\hat{F}_B$ of a magnetic field gradient (see also [13] [18]). The authors discovered the flux of spins in the following way: By an electric current through a point contact between ferromagnetic permalloy and paramagnetic aluminum electrons with aligned spin flow into the latter. There they diffuse and their spin direction is conserved for some nanoseconds. The difference of magnetization between aluminum and a second layer of permalloy induce an electric voltage of some picovolts between both metals. This can be measured with a precision voltmeter and had been called the 'inverse spin Hall effect' ($L_{qB}$ in table 1). With it in mind Johnson and Silsbee proposed the following system of three transport equations for a thermo-electro-magnetic system (cf. equ. (4)):

$$\boldsymbol{J}_Q = -L_{QT}\frac{\nabla T}{T^2} - L_{Q\phi}\frac{\nabla \phi}{T} + L_{QB}\frac{\nabla B}{T} \tag{7a}$$

$$\boldsymbol{J}_q = -L_{qT}\frac{\nabla T}{T^2} - L_{q\phi}\frac{\nabla \phi}{T} + L_{qB}\frac{\nabla B}{T} \tag{7b}$$

$$\boldsymbol{J}_M = -L_{MT}\frac{\nabla T}{T^2} - L_{M\phi}\frac{\nabla \phi}{T} + L_{MB}\frac{\nabla B}{T} \tag{7c}$$

Here one perceives two new reciprocity relations, namely

$$L_{QB} = -L_{MT} \quad \text{and} \quad L_{qB} = -L_{M\phi} \tag{8a,b}$$

(minus sign due to Casimir [7]).

After this discovery it took about ten years until one began to investigate the details of spin tranport. This was triggered by the development of new devices for electronic data procession. One had realized that writing and reading of magnetic bits works faster with spin transport than with induced currents. Using spins one obtains reading and writing times below 50 nanoseconds. After all spin transport needs fewer energy than induction switching. The spin Seebeck effect was discovered in 2008 ($L_{MT}$ [14]) and the spin Peltier effect ($L_{QB}$ [15]) six years later. In 2010 the term 'spincaloritronics' was established for the electrocaloric effect associated with a change of magnetization. Soon afterwards appeared a theoretical paper on 'magneto-mechano-thermal effects' with 16 kinetic coefficients and 6 new Onsager relations [30]. And it was followed by a publication on anisotropic crystals containing another 25 kinetic coefficients and 10 new Onsager relations [21]. What a mess of tasks for experimental physicists! Following these papers on the subject their number grew exponentially. This can be seen in a survey by Bauer, Saitoh and van Wees [23] with more than 100 relevant citations.

Unfortunately the study of literature is complicated by two facts: Nearly every author uses different designations for one and the same phenomenon. So for instance the spin Seebeck effect is called: giant magneto Seebeck, or spin dependent Seebeck, or tunnel magnetic Seebeck, or spin thermo-electric effect, or magneto-thermo force and so on. Moreover nearly every author uses other characters for one and the same property. So the reader finds himself in a jungle.

One further remark on notation is necessary: One uses the term (real or genuine) 'spin effect' if the transport is conducted by moving spins alone, i. e. by spin waves. No movement of electrons should be observed in this case as for instance in insulators. On the other hand one speaks of 'spin *dependent* effects' if spin transport results from the motion of electrons, atoms, phonons, polaritons etc. with their accompanying spins. Unfortunately this prescription is not obeyed by many authors. So one has to look carefully what had been measured or calculated.

## VIII. Experiments on spincaloritronics

Most of the recent experiments are carried out on thin films and at low temperatures [25]. This has two reasons: In most substances spin waves travel only some nanometers or micrometers wide at room temperature before they are dispersed or absorbed. Therefore one prefers thin layers which can be produced by evaporation in precise concentration and specified atomic structure. Moreover atomic arrangement and composition can be altered by suitable heat treatment in thin films. Because atomistic transport by electrons is slowed down mostly by phonon scattering it is promising to work at low temperatures where the phonon density is small.

As mentioned above spin currents are produced best by transport of polarized electrons through the boundary of a ferromagnetic and a paramagnetic substance. The transport can be driven either by a gradient of electric potential or of temperature. To detect the transported spins one can use different methods as the magneto-optic Kerr effect or the inverse spin Hall effect ([22] [24]) or by a diamant nitrogen vacancy magnetometer [29]. In Figure 4 is outlined the principle of a 'spin valve' working in this way. Similar microscopic devices are used in modern electronic data processing in order to obtain high density memories and short switching times. As mentioned above spintronic elements are favourable also because of their lower energy consumption compared to induction switching of bits.

## IX. Onsager relations in spincaloritronics

In 2014 appeared the first reliable measurement of Onsager relations in spin dependent Seebeck and Peltier effects [16]. 'Reliable' stands here for a study at one and the same sample with one and the same equipment. The results are shown in Figure 5 measured at room temperature. The sample consisted of three layers ($Ni_{80}Fe_{20}|Cu|Ni_{80}Fe_{20}$) of 15 nanometers thickness each. The results for $L_{qT}$ and $L_{Q\phi}$ (see Table 1) coincide within less than 5% below a Peltier current of 0.5 mA. At higher currents the linear region of equation (1) is exceeded and both coefficients diverge. Similar results had been published for a two-dimensional electron gas in a triplicate layer (In|Ga|InAs) at 240 mK [17].

In 2017 a summarizing survey [20] on the already known and the still to investigate effects of spincaloritronics contains 40 similar phenomena from 200 citations. But Onsager relations are touched upon with only one short remark in this paper. One may conclude that these relations are considered merely as an old hat. And this although they are so important for the judgement of microscopic reversibility and local equilibrium in the wanted devices. In 2006 already one has observed the Onsager relation between spin Hall effect and his counterpart, the elctric voltage caused by a spin current [24]. Another new observation is the 'first spin Ettingshausen Nernst effect' [31] (see Fig. 2): A longitudinal temperature gradient together with a perpendicular magnetic field produces a pure spin current in a platinum thin film. The reverse process and the corresponding Onsager relation is still awaiting ist discovery.

Since several years the spintronic community is very active hunting new phenomena. In 2017 the '8. International Workshop on Spincaloritronics' published the following statement: 'Today we know an enormous number of new phenomena with spin transport. Theory and experiment are encouraged to investigate these effects and make them available for application in electronic data processing' [26].

*To sum up: There is plenty of work to be done.*

## X. Appendix A. Transport equations

Since long ago we know the so called 'conventional' transport equations:

- $\boldsymbol{J}_q = -\hat{\sigma}\nabla\phi$  for electrical conduction, (a)
- $\boldsymbol{J}_Q = -\lambda\nabla T$  for thermal conduction, (b)
- $\boldsymbol{J}_N = -D\nabla c$  for particle conduction, (c)
- etc.

The units for fluxes $\boldsymbol{J}_X$ are defined by $[X]/(m^2s)$. Those for driving forces $\hat{\boldsymbol{F}}_Y$ are deduced from equation (2) as $J/([X]Km)$ using the unit of $\sigma$ $J/(Km^3s)$. $[X]$ is the unit of $X$.
*Attention:* The driving forces are *not* Netonian forces with the units of N (Newton).

The connection between conventional transport coeffisients $\hat{\sigma}$, $\lambda$, $D$ etc. and kinetic coefficients $L_{XY}$ can be obtained by equating the right sides of (1) and (a), (b), (c) etc. So one gets for instance $L_{QT} = \lambda\,T^2$, $L_q\phi = \hat{\sigma}\,T$ etc.

The names of many processes in Table 1 appear rather arbitrary throughout the literature. Some of them are not even specified, some have no name at all. The heat flux $\boldsymbol{J}_Q$ can be substituted by the entropy flux $\boldsymbol{J}_S$ if necessary but $\boldsymbol{J}_Q$ has become in use.

One can supplement Table 1 by still other pairs of fluxes and driving forces, for instance by strains and stresses, or by electric moments and electric field gradients, or by mass and gravitational forces etc. Last not least one has to consider the different directions of $\boldsymbol{J}$ and $\hat{\boldsymbol{F}}$ in crystalline solids and anisotropic liquids. In this case the kinetic coefficients become components of a tensor.

## XI. Appendix B. Short sketch of Onsager's derivation

It is not possible to reproduce Onsager's proof of his reciprocity relations in a few lines. But I shall try to explain the essential steps with some words: We begin with equation (4), $J_1 = L_{11}\hat{F}_1 + L_{12}\hat{F}_2$, $J_2 = L_{21}\hat{F}_1 + L_{22}\hat{F}_2$ etc., and we substitute for the flux $J$ a time-dependend term $\dot{a}$ of an extensive co-ordinate $a(t)$ describing the process in question. Next we substitue the driving force $\hat{F}$ by the derivative $\partial S/\partial a$ of the entropy $S$. This is based on equation (2) because $J \cdot \hat{F} = \sigma \propto S$. Now we look at the thermal fluctuations of $a$ around ist equilibrium value $a = 0$. Then we have from equation (4) $\dot{a}_j = \sum_k L_{jk}(\partial S/\partial a_k)$ with $j, k$ = 1, 2, 3 etc. We multiply this expression on both sides with an arbitrary term $a_r$ (trick 1) and we average both sides over time, on the right side with Boltzmann's probability $W(a)$ over $a(t)$ (trick 2). In this way $\partial S/\partial a_k$ can be replaced by $\partial W/\partial a_k$. Integrating this by parts over $da_j$ and approximating the boundaries in a clever way one obtains instead of (4) the expression $(1/\tau)\langle a_r \dot{a}_j \rangle = -k L_{jr}$ with the small time time interval $\tau$ and Boltzmann's constant $k$ (trick 3). Finally one assumes that the values of $(a + \tau)$ and $(a - \tau)$ are approximately equal on both sides of the maximum $a_{\max}(t)$ (trick 4, microscopic reversibility). Exchanging $(a + \tau)$ and $(a - \tau)$ in the last result one obtains after all the relation $L_{ik} - L_{ki} = 0$ q. e. d. To my knowledge the best detailed version of this deduction can be found in R. Becker's famous textbook *Theory of Heat* [34].

*Acknowledgement*: I am very obliged to my colleague Gerrit E. W. Bauer for critical reading the manuscript and for some very helpful suggestions.

# References

[1]  L. Onsager, Reciprocal Relatios in Irreversible Processes I and II, Phys. Rev. **37**, 405-426 and **38**, 2265-2279 (1931)

[2]  C. Kittel, *Elementary Statistical Physics*, Wiley, New York (1967)

[3]  H. B. Callen, *Thermodynamics*, Wiley, New York (1960)

[4]  F. Reif, *Fundamentals of Statistical and Thermal Physics*, McGraw-Hill, New York (1965)

[5]  L. B. Landau, E. M. Lifschitz, *Statistical Physics*, Elsevier, Amsterdam (1980)

[6]  S. R. de Groot, P. Masur, *Non-equilibrium Thermodynamics*, North-Holland, Amsterdam (1962)

[7]  H. B. G. Casimir, On Onsager's Principle of Microscopic Reversibility, Rev. Mod. Phys. **17**, 343-350 (1945)

[8]  L. E. Reichl, *A Modern Course in Statistical Physics*, Wiley, New York (1998)

[9]  F. Schwabl, *Statistical Mechanics*, Springer, Berlin (2006)

[10]  D. G. Miller, Thermodynamics of Irreversible Processes. The Experimantal Verification of Onsager Reciprocal Relations, Chem. Rev. **60**, 15-37 (1960)

[11]  R. L. Rowley, F. H. Horne, The Dufour Effect II. Experimental confirmation of the Onsager heat-mass reciprocal relation for a binary liquid mixture, J. Chem. Phys. **68**, 325-326 (1978)

[12]  M. Johnson, R. H. Silsbee, Thermodynamic analysis of interfacial transport and of the thermomagnetoelectric system, Phys. Rev. B **35**, 4959-4972 (1987)

[13]  K. Stierstadt, *Physik der Materie*, VCH, Weinheim (1989)

[14]  K. Uchida et al., Observation of the spin Seebeck effect, Nature **455**, 778-781 (2008)

[15]  J. Flipse et al., Observation of the Spin Peltier Effect for Magnetic Insulators, Phys. Rev Lett. **113**, 027601 (2014)

[16]  F. K. Dejene, J. Flipse, B. J. van Wees, Verification of the Thomson-Onsager reciprocity relation for spin caloritronics, arXiv 1408.2670v1 (2014)

[17]  J. Matthews et al., Experimental test of reciprocity relations in quantum thermoelectric transport, arXiv 1906.3694v2 (2014)

[18] A. D. Avery, B. L. Zink, Peltier Cooling and Onsager Reciprocity in Ferromagnetic Thin Films, Phys. Rev. Lett. **111**, 126602 (2013)

[19] G. Guevara-Carrion et al., Diffusion in Multicomponent Liquids: From Microscopic to Macroscopic Scales, J. Phys. Chem. B **120**, 12193-12210 (2016)

[20] H. Yu. S. D. Brechet, J.-P. Ansermet, Spin caloritronics, origin and outlook, Phys. Lett. A **381**, 825-837 (2017)

[21] P. Jacquod et al., Onsager relations in coupled electric, thermoelectric, and spin transport: The tenfold way, Phys. Rev. B **86**, 155118 (2012)

[22] J. E. Hirsch, Spin Hall Effect, Phys. Rev. Lett. **83**, 1834-1837 (1999)

[23] G. E. W. Bauer, E. Saitoh, B. J. van Wees, Spin caloritronics, Nature Mater. **11**, 391-399 (2012)

[24] E. Saitoh, M. Ueda, H. Miyajima, Conversion of spin current into charge current at room temperature: Inverse spin-Hall effect, Appl. Phys. Lett. **88**, 182509 (2006)

[25] A. Fert, Nobel Lecture: Origin, development, and future of spintronics, Rev. Mod. Phys. **80**, 1517-1530 (2008)

[26] J. Sinova, T. Jungwirth, Surprise from the spin Hall effect, Phys. Today **70**, Jul. 38-42 (2017)

[27] J. Schubert, *Physikalische Effekte*, Physik Verlag,Weinheim (1982)

[28] K. Stierstadt, *Thermodynamik*, Springer, Berlin (2010)

[29] C. Du et al., Control and local measurement of the spin chemical potential in a magnetic insulator, Science **357**, 195-198 (2017)

[30] G. E. W. Bauer et al., Nanoscale magnetic heat pumps and engines, Phys. Rev. B **81**, 024427 (2010)

[31] S. Meyer et al., Observation of the spin Nernst effect, Nature Mater. **16**, 977-981 (2017)

[32] M. Münzenberg, A. Thomas, Heiße Elektronik, Phys. i. u. Zeit **43**, 288-295 (2012)

[33] I. Žutić, J. Fabian, S. Das Sarma, Spintronics: Fundamentals and applications, Rev. Mod. Phys. **76**, 323-395 (2004)

[34] R. Becker, *Theory of Heat*, Springer, Berlin (1967)

# Figure captions

**Fig. 1.** General scheme for a transport process.

**Fig. 2.** Galvanomagnetic and thermomagnetic effects. The full lines are the causes, the broken lines the effects. $J_q$ and $J_Q$ are the electric and heat current density respectively. $B$ is the magnetic field, + and – the electric potential difference, $c$ (cold) and $w$ (warm) the temperature difference. No specific names are in use for the longitudinal effects.

**Fig. 3.** Experiments are much more time consuming and expensive than theories (cartoon by David Jackson, Univ. of Calif. at Berkeley).

**Fig. 4.** Principle set up of a spin valve. An electric current $I$ causes a spin flux from ferromagnet FM1 into the paramagnet PM. The flux extends partly as far as into the ferromagnet FM2. The spin density in PM and in FM2 can be detected by inverse spin Hall effect (ISHE, voltage $V$) or by the magneto-optical Kerr effect.

**Fig. 5.** Test of the Onsager relation for spin dependent Seebeck ($L_{qT}$) and Peltier ($L_{Q\phi}$) effects. Here the dependence of a voltage $V$ is plotted against the measured current $I$. $V$ is proportional to $L_{qT}$ (o) or to $L_{Q\phi}$ (∆) respectively. With increasing current the influence of nonlinear contributions to equation (1) is growing. The Onsager relation is perfectly obeyed for $I \rightarrow 0$.

# Table caption

**Tab. 1.** Linear transport processes.

Symbols: $T$ temperature, $P$ pressure, $\mu$ chemical potential, $\phi$ electrical potential, $B$ magnetic field, $Q$ heat energy, $V$ volume, $N$ particle number, $q$ electric charge, $M$ magnetic moment, $\lambda$ heat conductivity, $\eta$ shear viscosity, $D$ diffusion constant, $\hat{\sigma}$ electrical conductivity, $D_M$ spin diffusion constant.

The so called conjugated effects are indicated in bold in the second diagonal.

The flux $\boldsymbol{J}_m$ of mass can be obtained by multiplying $\boldsymbol{J}_V$ with the mass density $\rho$. The conjugated driving force is $\hat{\boldsymbol{F}}_\rho = \hat{\boldsymbol{F}}_P / \rho = -\nabla P / (\rho T)$.

The flux of entropy can be obtained by dividing $\boldsymbol{J}_Q$ by the temperature $T$. The conjugated driving force is $\hat{\boldsymbol{F}}_S = T\hat{\boldsymbol{F}}_T = -\Delta T / T$.

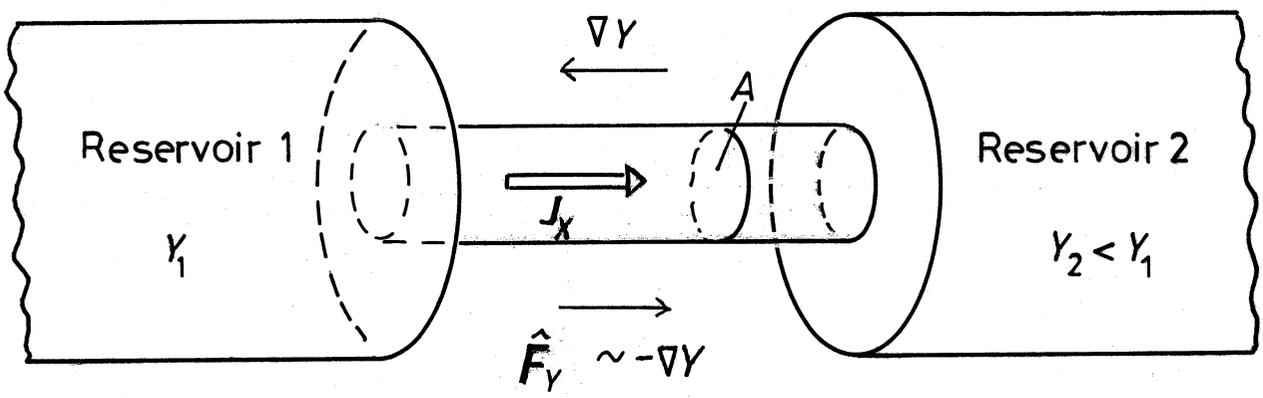

Fig. 1

## Galvanomagnetism

| transversal | | longitudinal |
|---|---|---|
| Hall-effect | Thomson-effect | →B<br>→$J_q$<br>+ − |
| Ettingshausen-effect | Nernst-effect | →B<br>→$J_q$<br>c   w |

## Thermomagnetism

| transversal | | longitudinal |
|---|---|---|
| 1. Righi-Leduc-effect | 2. Righi-Leduc-effect | →B<br>→$J_Q$<br>w   c |
| 1. Ettingshausen-Nernst-effect | 2. Ettingshausen-Nernst-effect | →B<br>→$J_Q$<br>−   + |

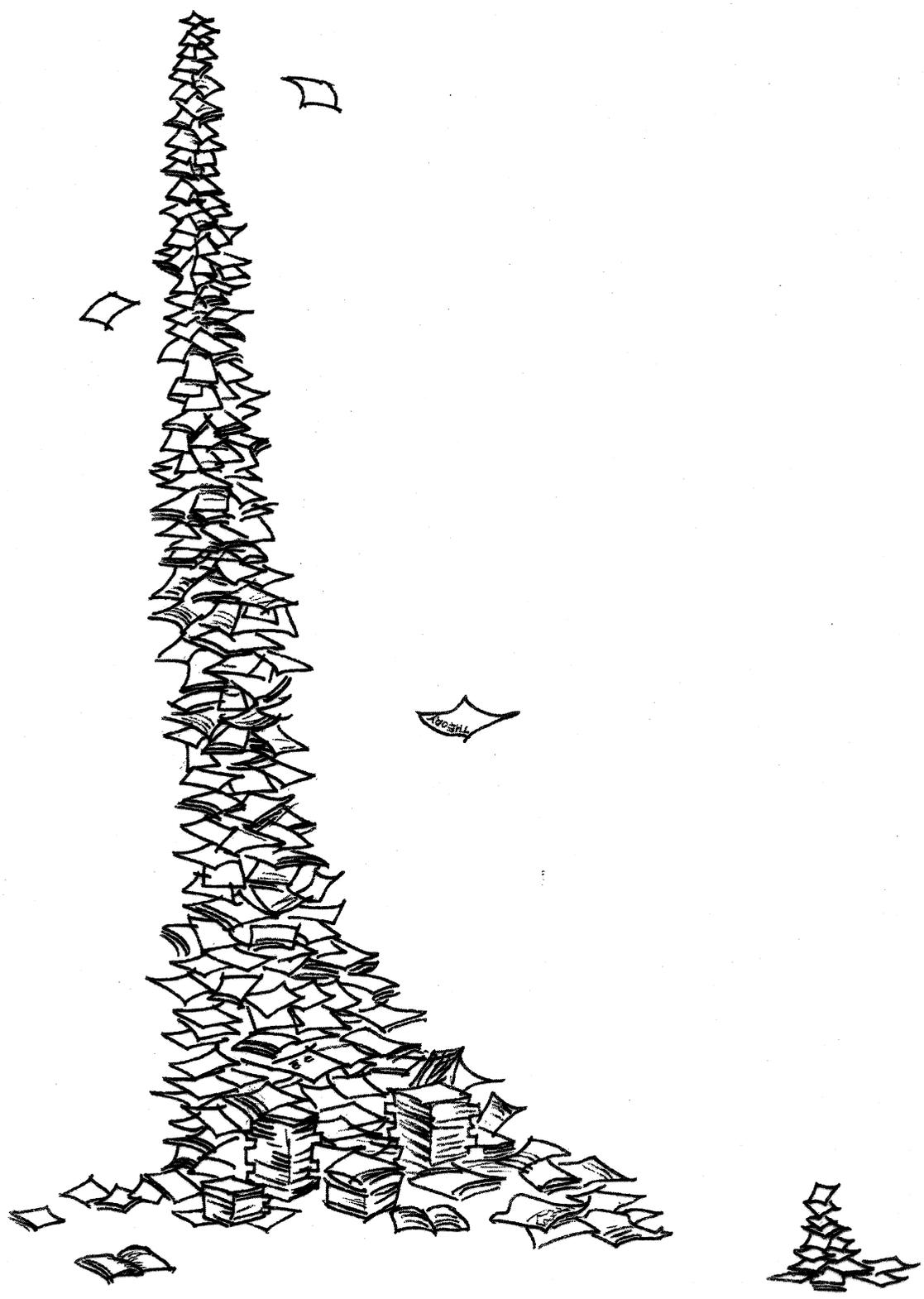

THEORY   EXPERIMENT

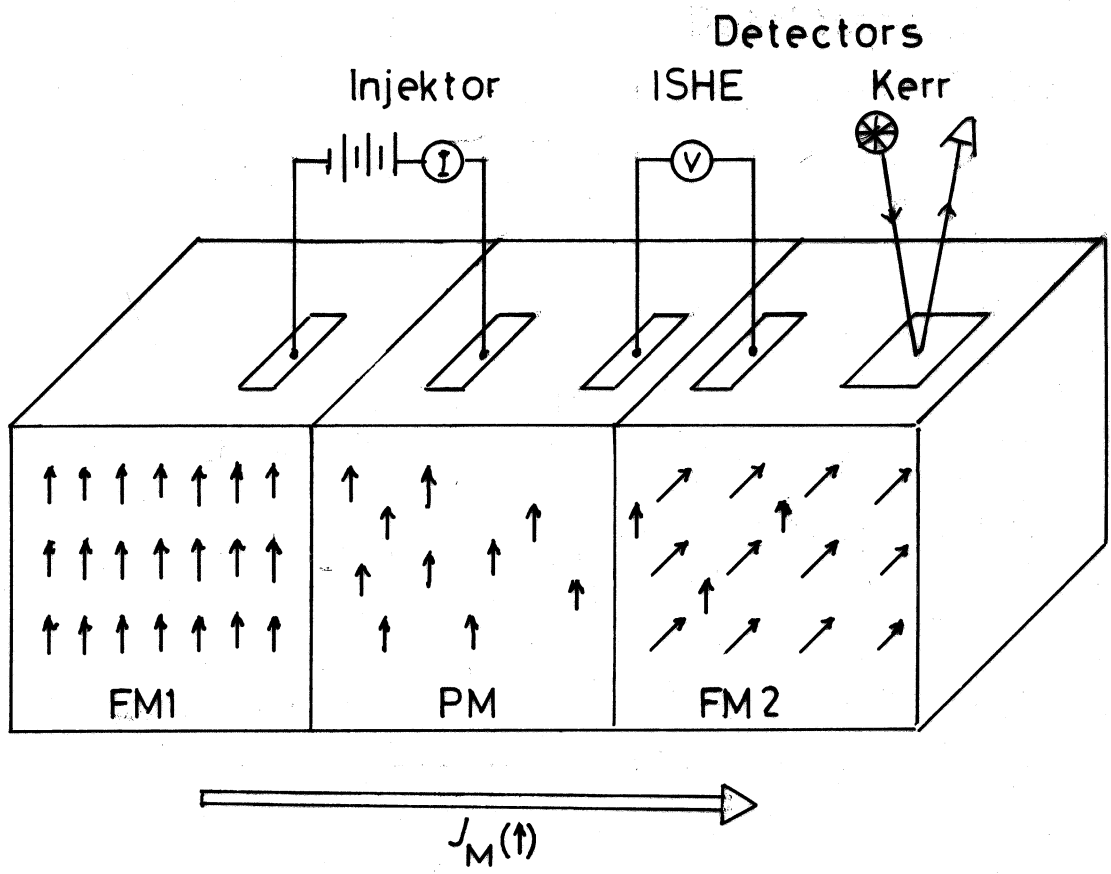

Fig. 4

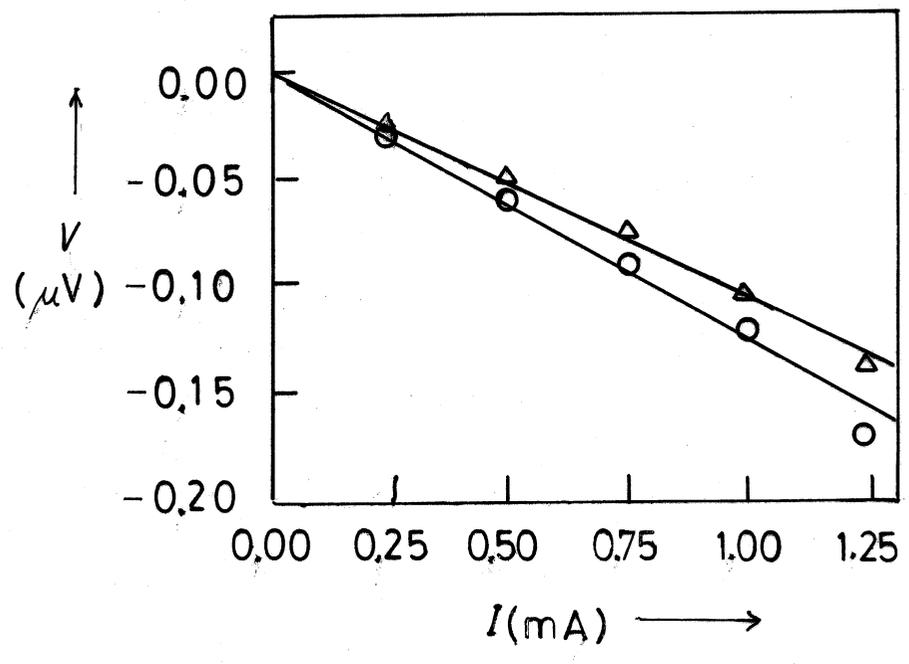

Fig. 5

**Table 1  Linear transport processes.**

| $\hat{F}_Y$: <br> $J_X$: | $\hat{F}_T = \nabla\left(\dfrac{1}{T}\right) = -\dfrac{\nabla T}{T^2}$ | $\hat{F}_P = -\dfrac{\nabla P}{T}$ | $\hat{F}_\mu = -\dfrac{\nabla \mu}{T}$ | $\hat{F}_\phi = -\dfrac{\nabla \phi}{T} = \dfrac{E}{T}$ | $\hat{F}_B = \dfrac{\nabla B}{T}$ |
|---|---|---|---|---|---|
| $J_Q$ <br> (Jm$^{-2}$s$^{-1}$) | **Heat conduction** <br><br> $L_{QT} \sim \lambda$ | Pressure caloric effect <br> (Mechano caloric effect) <br> $L_{QP}$ | Diffusion caloric effect <br> (Dufour effect) <br> $L_{Q\mu}$ | Electro caloric effect <br> (Peltier effect) <br> $L_{Q\phi}$ | Magneto caloric effect <br> (Spin Peltier effect) <br> $L_{QB}$ |
| $J_V$ <br> (ms$^{-1}$) | Thermo volume flux <br> (Thermo mechanic effect) <br> $L_{VT}$ | **Volume flux** <br> (Pressure balance) <br><br> $L_{VP} \sim 1/\eta$ | Diffusion volume effect <br><br> $L_{V\mu}$ | Electrokinesis <br><br> $L_{V\phi}$ | Magnetostriction <br><br> $L_{VB}$ |
| $J_N$ <br> (m$^{-2}$s$^{-1}$) | Thermodiffusion <br> (Soret effect) <br> $L_{NT}$ | Pressure osmosis <br><br> $L_{NP}$ | **Diffusion** <br><br> $L_{N\mu} \sim D$ | Electroosmosis <br><br> $L_{N\phi}$ | Magneto diffusion <br><br> $L_{NB}$ |
| $J_q$ <br> (Am$^{-2}$) | Thermo current <br> (Seebeck effect) <br> $L_{qT}$ | Pressure electric effect <br> (Streaming current) <br> $L_{qP}$ | Diffusion electric effect <br><br> $L_{q\mu}$ | **Electric conduction** <br><br> $L_{q\phi} \sim \hat{\sigma}$ | Magneto electric effect <br> (Inverse spin Hall effect) <br> $L_{qB}$ |
| $J_M$ <br> (As$^{-1}$) | Thermo spin current <br> (Spin Seebeck effect) <br> $L_{MT}$ | Pressure magnetic effect <br><br> $L_{MP}$ | Spin diffusion <br><br> $L_{M\mu}$ | Electro magnetic effect <br> (Spin Hall effect) <br> $L_{M\phi}$ | **Spin drift** <br><br> $L_{MB} \sim D_M$ |